\documentclass[a4paper,11pt]{article}
\pdfoutput=1 

\usepackage{jcappub} 

\usepackage[T1]{fontenc} 
\usepackage{url}
\usepackage{blindtext}
\usepackage{hyperref}
\usepackage{amsmath}
\usepackage{siunitx}

\title{\boldmath The impact of gravitational lensing in the reconstruction of stellar orbits around Sgr A*
}

\author[a,b]{S. Pietroni,
}
\author[a,b]{V. Bozza}

\affiliation[a]{Dipartimento di Fisica “E.R. Caianiello”, Università di Salerno, Via Giovanni Paolo II 132,
I-84084 Fisciano, Italy}
\affiliation[b]{ Istituto Nazionale di Fisica Nucleare, Sezione di Napoli, Via Cintia, 80126, Napoli, Italy}

\emailAdd{spietroni@unisa.it}
\emailAdd{silviapietroni@hotmail.com}
\emailAdd{valboz@sa.infn.it}

\abstract{After the amazing discoveries by the GRAVITY collaboration in the last few years on the star S2 orbiting the black hole Sgr A* in the center of the Milky Way, we present a detailed investigation of the impact of gravitational lensing on the reconstruction of stellar orbits around this massive black hole.
We evaluate the lensing astrometric effects on the stars S2, S38 and S55 and how these systematically affect the derived orbital parameters. The effect is below current uncertainties, but not negligible. With the addition of more observations on these stars, it will be possible to let the astrometric shift by lensing emerge from the statistical noise and be finally detected.

By repeating the analysis on a smaller semimajor axis $a$ and various inclinations $i$, we are able to quantify the lensing effects on a broader range of parameters. As expected, for smaller semimajor axes and for nearly edge-on orbits lensing effects increase by about an order of magnitude.
}

\begin{document}
\maketitle
\flushbottom

\section{Introduction}
\label{sec:intro}

The Galactic Center (GC) is one of the most attractive regions of the Milky Way that has been monitored for about three decades \cite{2019Msngr.178...26G}.
The Sgr A complex of radio sources extends over the central 10 parsecs of our Galaxy. A compact object called Sgr A* with a mass of several million solar masses lies in its center, traditionally accepted as a massive black hole (MBH). 
\cite{gillessen2009orbit,gillessen2017update}.

The central parsec hosts old, late-type
red giants, supergiants, asymptotic giant branch stars, but also many hot young, early-type
stars, like post-main sequence blue supergiants and Wolf-Rayet stars \cite{2013CQGra..30x4003F,2003ApJ...586L.127G,2005ApJ...628..246E,2009ApJ...692.1075G,2017ApJ...837...30G,2012Sci...338...84M,2016ApJ...830...17B,2016ApJ...821...44F,2002Natur.419..694S,2012RAA....12..995M,2009A&A...502...91S}, and two different observational campaigns from VLT and Keck observatories, revealed a hundred of B stars within $1$ arcsecond in the neighborhood of Sgr A*, known as the S-stars \cite{2005ApJ...628..246E,2017ApJ...837...30G,2021A&A...645A.127G,2012Sci...338...84M,2016ApJ...830...17B}. Many of the orbits of these stars have been reconstructed in detail after many years of observations.

A major breakthrough in the astrometric precision in these campaigns has been achieved when GRAVITY became operational. GRAVITY is an interferometer part of the second generation of VLTI \cite{2008SPIE.7013E..2AE,2019hsax.conf..609A,2021A&A...647A..59G,2017A&A...602A..94G}, the Very Large Telescope Interferometer, which combines the light of the four VLT telescopes (the four Unit Telescopes (UTs)) and the Auxiliary Telescopes (ATs) operated by the European Southern Observatory on Cerro Paranal in the Atacama Desert of northern Chile. GRAVITY provides a spectro-interferometry in the K-band between 2 and 2.4 $\mu$m and a resolution between 4 mas (milli-arcsecond) and 50 mas with the UTs, and between 2 mas and 140 mas with the ATs.
GRAVITY observed Sgr A* for 6 months in 2019 in order to detect the weakest objects closer to it and follow their movement across the sky. 


Indeed, the major result of these observations has been the detection of the relativistic precession of the star S2  \cite{2020A&A...636L...5G}, which has marked an exceptional milestone for General Relativity (GR), astrophysical black holes and Near Infrared interferometric observations. S2 is characterized by the most favorable characteristics for such gravity test: it is bright enough for clear detection both in astrometry and spectroscopy, it has a relatively small orbital period of about 16 years, it comes as close as 120 au to the central black hole. Its nearly elliptical orbit only perturbed by GR effects is evidence of the existence of a massive compact object located at its focus.

Astrometric measurements of the positions of the stars around a black hole are obviously subject to light bending, which shifts the observed light outwards from the black hole. There is a number of expectations related to gravitational lensing by Sgr A*. The apparent density of background stars should decrease around the black hole  \cite{1992ApJ...387L..65W}. The orbits of stars and the velocity distribution should be distorted \cite{1998AcA....48..413J}. Extremely dim negative parity images should appear on the opposite side of the stars with respect to the black hole \cite{2003A&A...409..809D,2004ApJ...611.1045B,2005ApJ...627..790B}. These will be probably detectable by next generation instruments, such as the ELT \cite{2021ApJ...915L..33M}. However, the most accessible effect of light bending remains the astrometric shift of the primary image, which should be within reach of GRAVITY for a number of S-stars \cite{2012ApJ...753...56B}. The deflection is generally too weak for known stars to bear any memory of the spin of the black hole \cite{2012ApJ...753...56B,2015ApJ...809..127Z}. Other GR effects such as the relativistic precession or time delay tend to dominate \cite{2017A&A...608A..60G} and return more information on the characteristics of the black hole, including its angular momentum \cite{2018MNRAS.476.3600W}. 

Nevertheless, even a very weak distortion of the orbits of S-stars introduces systematic errors in the determination of the orbital parameters if the astrometric shift is neglected. This error needs to be correctly quantified and compared to statistical uncertainties. The purpose of the present paper is to assess the level of systematics induced by gravitational lensing in the orbits of three short period S-stars (S2, S38, S55). This is achieved by forcing an orbital fit without lensing to reproduce the orbit including lensing and evaluating the shift in the parameters. As expected, we find that the systematic error is below current uncertainties but not so far. After that, with the same spirit of similar studies \cite{2018MNRAS.476.3600W}, by varying the semimajor axis and the orbital inclination of the star S2, we determine the region of the orbital parameters in which a similar star would show evidence of gravitational lensing beyond statistical noise.

The paper is organized as follows. Section 2 introduces the three stars investigated in our study and their orbits. Section 3 discusses gravitational lensing by Sgr A*. Section 4 evaluates the impact of gravitational lensing on the aforementioned stars. Section 5 repeats the same study on a clone of star S2 with different semimajor axis and inclination. Section 6 contains the conclusions.

\section{The stars S2, S38 and S55}\label{kep}

The S-stars present a variety of orbital parameters, according to an isotropic distribution with projected density $\Sigma(\theta) \sim \theta^{-0.3}$ \cite{2018A&A...609A..28B}. Many S-stars lie on very eccentric long-period orbits. The most favorable case for a full orbital reconstruction is provided by low-period stars, with $P\leq 20$ yrs, which are suitable for a long-term follow-up. The three stars discussed in this paper fulfill this requirement. 

S2 (or S02 in the UCLA nomenclature) is a single and slowly rotating  main-sequence B-star of age $\approx6$ Myr, apparent magnitude $K=13.95$ and estimated mass $M_{S2}=19.5\times10^6$ $M_\odot$ \cite{2020A&A...636L...5G,2008ApJ...672L.119M,2010MNRAS.409.1146G,2017A&A...602A..94G,2004PhyA..332...89C,2017ApJ...847..120H}. S2 represents the only case in which a prograde orbital precession of 12 arcminutes has been measured. Therefore, here we refer to the osculating orbital parameters.

In Table \ref{tab:orpar} we display the current estimates of S2 orbital parameters as in Refs. \cite{2020A&A...636L...5G,2022A&A...657A..82G}, where $a$ is the semimajor axis of the ellipse, $e$ its eccentricity, $i$ is the angle of inclination between the real orbit and the observation plane, $\Omega$ is the angle of the ascending node measured from the north direction, $\omega$ is the argument of the pericenter, $T_0$ is the epoch of pericenter passage, $P$ is the orbital period, $M_{BH}$ the black hole mass, 
$D_\bullet$ is the derived distance to the Galactic center as shown in Table \ref{mbh}.

\begin{table}[t]
\centering
\begin{tabular}{|c|c|c|c|c|}
\hline
\multicolumn{5}{|c|}{\textbf{Osculating Orbital Parameters}}\\
\hline
\textbf{Parameter} & \textbf{Unit} & \textbf{S2}  & \textbf{S38}   & \textbf{S55}\\
\hline
$a$ & mas &  $124.95\pm0.04$ & $142.54\pm0.04$ & $104.40\pm0.05$\\
$e$ &  &  $0.88441\pm0.00006$ & $0.81451\pm0.00015$   & $0.72669\pm0.00020$ \\
$i$  & $^{\circ}$  & $134.70\pm0.03$ & $166.65\pm0.4$ &  $158.52\pm0.22$ \\
$\omega$  & $^{\circ}$  & $66.25\pm0.03$ & $27.17\pm1.02$ & $322.78\pm1.13$\\
$\Omega$   & $^{\circ}$ &  $228.19\pm0.03$  & $109.45\pm1$  &  $314.94\pm1.14$  \\
$T_0$  &  Years  &  $2018.38\pm0.00$  & $2003.15\pm0.01$ & $2021.69\pm0.01$ \\
$P$   &  Years  & $16.046\pm0.001$  &   $19.55\pm0.01$ & $12.25\pm0.01$ \\
\hline

\end{tabular}
\caption{\label{tab:orpar} S2, S38 and S35 osculating orbital parameters from Gravity Collaboration's fit   \cite{2022A&A...657A..82G,2020A&A...636L...5G}.}
\end{table}

\begin{table}[]
    \centering
    
    \begin{tabular}{|c|c|c|}
    \hline
\textbf{Parameter} & \textbf{Unit}  & \textbf{Value}\\
\hline
$M_{BH}$  & $10^6$ $M_\odot$  & $4.33\pm0.01$ \\
\hline
$D_\bullet$  & pc  & $8246.7\pm9.3$ \\
\hline
    \end{tabular}
    \caption{The black hole mass and the derived distance to the Galactic Center \cite{2022A&A...657A..82G,2020A&A...636L...5G}.}
    \label{mbh}
\end{table}

The relativistic precession changes the orbit orientation by increasing $\omega$ by 12' per orbit. The last apocenter passage was in 2010, the next one is expected in 2026, the furthest distance from the MBH is $r_{max}=1954.24$ au according to Eq.\eqref{apo}, the last pericenter passage is $T_0=2018.38$ (May 19th 2018) \cite{2009ApJ...707L.114G,2018A&A...618L..10G}, the closest distance estimated according to Eq.\eqref{peri} is $r_{min}=119.87$ au and the next pericenter passage is expected to be in 2034.


S38 is a main-sequence B-star of apparent magnitude $K=17$. In Table \ref{tab:orpar} we display the orbital parameters for S38 as in Ref. \cite{2020A&A...636L...5G}. 
The last apocenter passage was at the end of 2012, the next one is expected in 2032, the furthest distance from the MBH is $r_{max}=2146.66$ au according to Eq.\eqref{apo}, the last pericenter passage is $T_0=2003.15$, the closest distance estimated according to Eq.\eqref{peri} is $r_{min}=219.44$ au and the next pericenter passage is expected to be in 2022.


S55 is a main-sequence B-star of apparent magnitude $K=17.5$.  In Table \ref{tab:orpar} we display the orbital parameters for S55 as in Ref. \cite{2020A&A...636L...5G}.
The last apocenter passage was in 2015, the next one is expected in 2027, the furthest distance from the MBH is $r_{max}=1496.17$ au according to Eq.\eqref{apo}, the last pericenter passage is $T_0=2021.69$, the closest distance estimated according to Eq.\eqref{peri} is $r_{min}=236.82$ au and the next pericenter passage is expected to be at the end of 2033.


For the orbit calculation of the S-stars, we start from the Lagrangian of a particle in the Schwarzschild metric

\begin{equation}\mathcal{L}=\frac{1}{2
}m\biggl[A(r)c^2\dot{t}
^2-A(r)^{-1}\dot{r}
^2-r
^2\dot{\phi}^2\biggr]\label{2.14}\end{equation}

where $m$ is the reduced mass of the star, $\phi$ is the true anomaly, $r$ is the radial coordinate, $t$ is the time for the observer at infinity and dots denote derivatives with respect to the proper time $\tau$. The Schwarzschild metric is characterized by
\begin{equation}
    A(r)=\biggl(1-\frac{R_S}{r}
\biggr)
\end{equation}
where
\begin{equation}R_S=\frac{2GM_{BH}}{c^2}\end{equation}
is the Schwarzschild radius, $G$ is the gravitational constant, $c$ is the speed of light.

The equations of motion can be expressed in terms of the constants of motion $E$ and $L$ as

\begin{equation}\dot{t}
=\frac{E}{
c^2 \biggl(1-\frac{R_S}{r}
\biggr)}\label{tpunto}\end{equation}



\begin{equation}\dot{\phi}
=\frac{L}{
r^2}\label{fipunto}\end{equation}

\begin{equation}\dot{r}
=\pm\sqrt{ \frac{E^2}{ c^2
}-\biggl(1-\frac{R_S}{r}
\biggr)\biggl(1+\frac{ L^2}{
r
^2}\biggr)\label{2.29}}\end{equation}
where the negative solution is for the particle approaching the black hole, the positive solution is for the particle moving away from the black hole. $E$ represents the energy and $L$ is the angular momentum of the particle per unit mass which we can get by solving the system with Eq.\eqref{2.29} evaluated for $r=r_{max}$ and $r=r_{min}$, where
\begin{equation}r_{max}=a (1+e)\label{apo}\end{equation}
\begin{equation}r_{min}=a (1-e)\label{peri}\end{equation}
are the apocenter and the pericenter respectively as functions of the semimajor axis and the eccentricity.

The equation of motion is \cite{BecerraVergara2020}:

\begin{equation}\ddot{r}
=-\frac{A(r)}{2}\biggl(\frac{dA(r)}{dr}\dot{t}
-\frac{1}{A(r)^2}\frac{dA(r)}{dr}\dot{r}
-2r\dot{\phi}
^2\biggr).\label{eqmo}\end{equation}

We solve Eq.\eqref{eqmo} for $r=r(\tau)$ numerically from the last pericenter passage $T_0$ starting from null velocity as initial conditions, Eq.\eqref{fipunto} and Eq.\eqref{tpunto} respectively give $\phi=\phi(\tau)$ and $t=t(\tau)$, then we invert the latter in order to have $\tau=\tau(t)$ to finally obtain $r=r(t)$ and $\phi=\phi(t)$ which give the position and the true anomaly as functions of the observer time. All these solutions are found numerically and their stability over several periods has been carefully checked. We have then evaluated the orbital precessions of our three stars by comparing the true anomaly at two consecutive pericenters, i.e. minima of $r$.
For the relativistic precession of S2 we find $\delta\phi=\phi(P)-2\pi=12'.17$, in agreement with GRAVITY results \cite{2020A&A...636L...5G}. For S38 we get $\delta\phi=\phi(P)-2\pi=6'.79$ and for S55 $\delta\phi=\phi(P)-2\pi=6'.46$.

Finally, the orientation of the orbits in space is described by the standard orbital elements $i$, $\Omega$, $\omega$. The detailed geometric transformation from $r(t)$ and $\phi(t)$ to the position in the sky described by $\Delta R.A.$ and $\Delta Dec.$ is reported in appendix \ref{orbpar} for completeness.

\begin{figure}[!htbp]
    \centering
    \includegraphics[width=7cm]{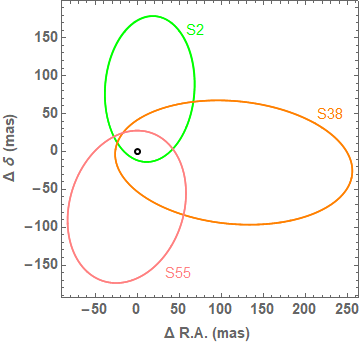}
    \caption{S2, S38 and S55 orbit reconstruction around Sgr A*.}
    \label{all3}
\end{figure}

The orbits of the three stars are shown in the left panel of Fig. \ref{all3} as they appear on the observer sky. The right panel shows them together with a selection of other potentially interesting S-stars that have been detected.

\section{Gravitational lensing by Sgr A*}

The light from the S-stars is deviated by the gravitational field of Sgr A*. The deflection is generally weak for the primary image, save for sources very close to the black hole and well-aligned behind it. However, this is precisely the situation in which gravitational lensing effects become appreciable. Therefore, we decide to keep the calculation of the deflection exact so as not to miss any potentially interesting effects. The source position is tracked by its radial coordinate $r_S\equiv r(t)$ as calculated from the orbital motion, and by the angle $\gamma$ between the source position and the optical axis, defined as the line of sight from the observer to the black hole. In terms of the true anomaly and the orbital elements defined before, this angle is

\begin{equation}\gamma(t)=\arccos{[\sin{(\phi(t)+\omega)}\sin{i}]}
\end{equation}
and ranges from 0 (perfect
alignment of the source behind the lens) to $\pi$ (perfect anti-alignment, with the source in front of
the lens).

A photon travelling from some radial coordinate $r_1$ to $r_2$ experiences an azimutal shift
\begin{equation}
    I(r_1,r_2;r_0)= \int_{r_1}^{r_2}\frac{dr}{r\sqrt{\frac{r^2}{r_0^2}\biggl(1-\frac{R_S}{r_0}
\biggr)-\biggl(1-\frac{R_S}{r}
\biggr)}},
\end{equation}
where $r_0$ is the closest approach distance of the null geodesic on which the photon is travelling. It is related to the impact parameter $u$ of the asymptotic trajectory at infinity as

\begin{equation}u=\frac{r_0}{\sqrt{1-\frac{R_S}{r_0}.
}}\label{J}\end{equation}

For a photon emitted by a source at $r_S$ and reaching an observer at distance $D_\bullet$, we have two possibilities: the photon may approach the black hole, reach the minimum distance $r_0$ and then travel toward the observer; otherwise, the photon may directly travel from $r_S$ to the observer at $D_\bullet$ without approaching the black hole. This is the case when the source is in front of the black hole as seen by the observer. Therefore, the azimutal shift is 

\begin{equation}
\Delta\phi(r_S,r_0)=\begin{cases}
I(r_0,r_S;r_0)+I(r_0,D_\bullet;r_0) \; \; \mathrm{if} \; \gamma< \gamma_0 \\
I(r_S,D_\bullet;r_0) \; \; \mathrm{if} \; \gamma\geq \gamma_0
\end{cases}
\label{az}
\end{equation}

The angle $\gamma_0$ distinguishing the two cases corresponds to the limiting case in which the photon is emitted with $r_0=r_S$. It can be calculated for a given source distance by imposing 
\begin{equation}
    I(r_S,D_\bullet;r_S)=\pi - \gamma_0,
\end{equation}
but is obviously very close to $\pi/2$ in all practical situations.

Finally, by imposing that the azimutal shift is just the correct one to let the photon move from $\gamma$ to $\pi$ (where the observer lies), we have a lens equation
\begin{equation}
    \Delta \phi(r_S,r_0)=\pi - \gamma.
\end{equation}

We solve numerically for the closest approach $r_0$ and find the corresponding impact parameter $u$. The angular distance of the image from the black hole as seen by the observer is then 
\begin{equation}\theta=\frac{u}{D_\bullet},\label{imp}
\end{equation}
since the observer distance $D_\bullet$ is much larger than any $u$ of our interest.

The star position including the astrometric shift is then obtained by increasing the angular coordinates $\Delta Dec.$ and $\Delta R.A.$ so that the angular distance from the black hole matches $\theta$ while keeping the position angle fixed. Note that the astrometric shift steadily increases from zero (when the source is in front of the black hole with $\gamma=\pi$) to the maximum value corresponding to the Einstein radius $\theta_E=\sqrt{2R_S \; r_S/D_\bullet(D_\bullet+r_S)}$, which is reached if the source is behind the black hole with $\gamma=0$.

As already stated before, we neglect any lensing effects from the spin of the black hole following the findings of previous works \cite{2012ApJ...753...56B,2015ApJ...809..127Z}.


\section{Impact of gravitational lensing on the reconstruction of the orbits of S- stars}
\label{lm}

The orbital fit of the S-stars is typically performed ignoring the astrometric shift due to gravitational lensing \cite{2017A&A...608A..60G,2019Sci...365..664D}. However, forcing an orbital fit without including gravitational lensing may lead to a systematic error in the derived parameters. The most obvious one is the semimajor axis: since the astrometric shift pushes the observed images away from the black hole, the real semimajor axis should be slightly smaller than what we naively deduce from the observations. However, as we shall see, the 3-D reconstruction of the orbital parameters may lead to some combined effects on different correlated parameters.

In order to quantify this systematic error, we create simulated observations of the S-stars along their orbits including the astrometric shift due to gravitational lensing. These ``data points'' are generated from the tabulated orbital elements as reported by recent observations \cite{2022A&A...657A..82G} and summarized in Table \ref{tab:orpar}. We then try to fit these data points with orbits that do not include gravitational lensing. The new orbital elements will be slightly offset from the original ones and quantify the systematic error committed by ignoring gravitational lensing.

Since the offsets from the original parameters are very small, it is sufficient to perform a few iterations with the Gauss-Newton method to find the new minimum of the $\chi^2$ \cite{madsen2004methods}. In detail, denoting the 7 orbital elements with the vector
\begin{equation}
\boldsymbol{\eta}=(a,e,i,\omega,\Omega,T_0,P),
\end{equation}
and the shift in the parameters with

\begin{equation}\boldsymbol{\delta }=(\delta_a,\delta _e,\delta_i,\delta_\omega,\delta_\Omega,\delta_{T_0},\delta_P),
    \end{equation}
we compare the positions including lensing $f_{lens,i}(\boldsymbol{\eta})$ and the positions without lensing $f_{0,i}(\boldsymbol{\eta})$. Here the index $i$ spans the number of generated data points $n$, including both $\Delta Dec.$ and $\Delta R.A.$. Introducing the gradients

\begin{equation}{\displaystyle \mathbf {J} _{i}={\frac {\partial f_{0,i}(\boldsymbol{\eta})}{\partial {\boldsymbol {\eta }}}}}
\end{equation}
the shifts are obtained by solving the linear equation


\begin{equation}{\displaystyle \left(\mathbf {J} ^{\mathrm {T} }\mathbf {J} \right){\boldsymbol {\delta }}=\mathbf {J} ^{\mathrm {T} }\left[\mathbf{f_{lens}} -\mathbf {f_0} \left({\boldsymbol {\eta }}\right)\right]}\end{equation}

where ${\displaystyle \mathbf {J} }$  is the Jacobian matrix, whose ${\displaystyle i}$-th row equals ${\displaystyle \mathbf {J} _{i}}$, and where ${\displaystyle \mathbf {f_0} \left({\boldsymbol {\eta }}\right)}$ and $\mathbf{f_{lens}}$  are vectors with ${\displaystyle i}$-th component $f_{0,i}$ and $ f_{lens,i}$ respectively. The calculation of the offsets $\boldsymbol{\delta}$ can be iterated so as to minimize the $\chi^2$

\begin{equation}\chi^2=\sum _{i=1}^{n}\left[\frac{f_{lens,i}-f_{0,i}(\boldsymbol{\eta}+\boldsymbol{\delta})]}{\sigma_i}\right]^{2},  \end{equation}
where $\sigma_i$ is the statistical uncertainty in each measurement. Here for simplicity we adopt a standard value of $\sigma=0.4$ mas for the uncertainty of each individual measurement, as reported by GRAVITY \cite{2022A&A...657A..82G}. A different value has no effect in the estimate of the systematic error calculated with our procedure, but is important to assess the relevance of systematic errors compared to statistical errors. In fact, the statistical uncertainty in each parameter scales as $\sigma/\sqrt{n}$. 

\begin{figure}[t]
\centering
   \includegraphics[width=7.5cm]{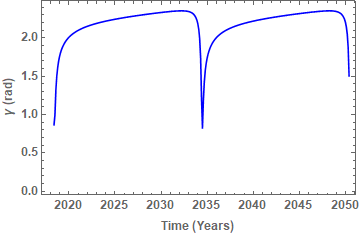}
    \includegraphics[width=7.5cm]{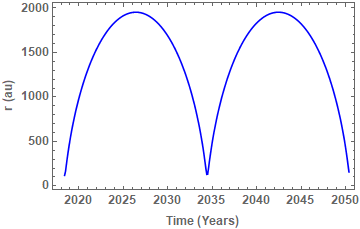}
    \includegraphics[width=7.5cm]{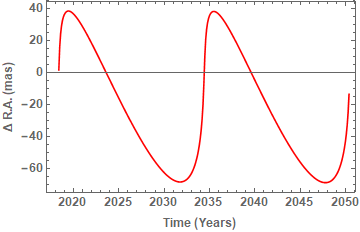}
    \includegraphics[width=7.5cm]{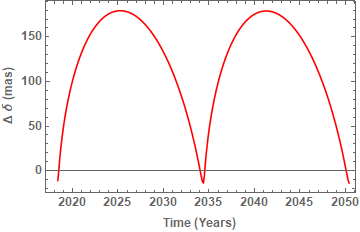}\\
    \includegraphics[width=7.5cm]{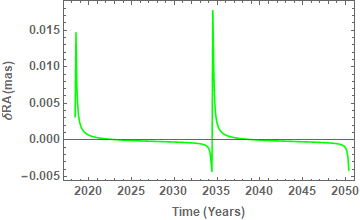}
     \includegraphics[width=7.5cm]{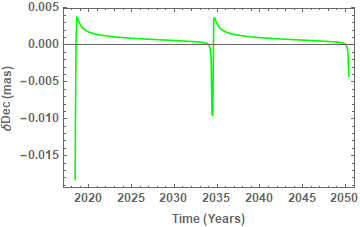}\\
     \caption{Analysis for S2. First row: $\gamma(t)$ (top left) and $r(t)$ (top right)
     as functions of time; 
     second row: right ascension (on the left) and declination (on the right); last row: astrometric shifts in right ascension (left) and declination (on the right). Computations are done starting from $T_0$ over two periods. 
     }
    \label{extra}
\end{figure}

\begin{figure}[!htbp]
\centering
   \includegraphics[width=7.5cm]{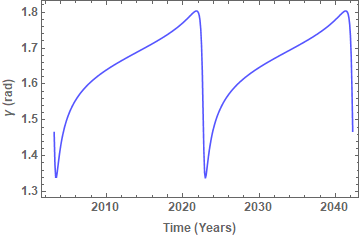}
     \includegraphics[width=7.5cm]{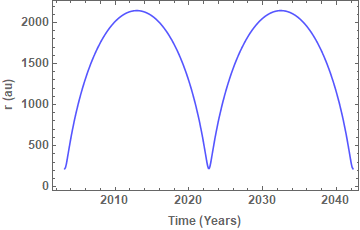}
    \includegraphics[width=7.5cm]{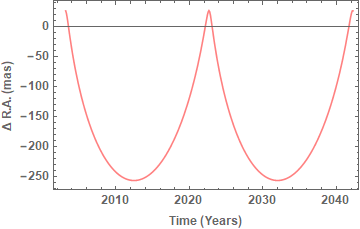}
    \includegraphics[width=7.5cm]{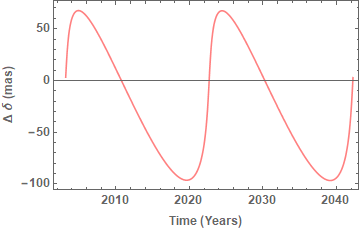}\\
    \includegraphics[width=7.5cm]{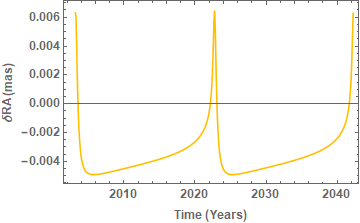}
     \includegraphics[width=7.5cm]{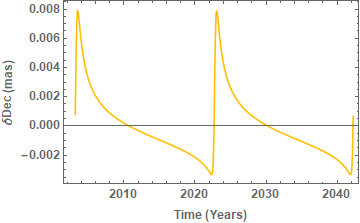}\\
     \caption{Analysis for S38. First row: $\gamma(t)$ (top left) and $r(t)$ (top right)
     as functions of time; 
     second row: right ascension (on the left) and declination (on the right); last row: astrometric shifts in right ascension (left) and declination (on the right). Computations are done starting from $T_0$ over two periods.
     }
    \label{extra38}
\end{figure}

\begin{figure}[!htbp]
\centering
   \includegraphics[width=7.5cm]{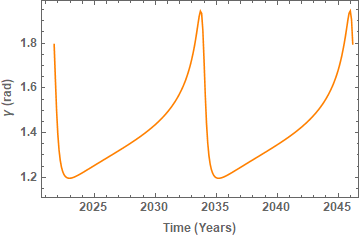}
     \includegraphics[width=7.5cm]{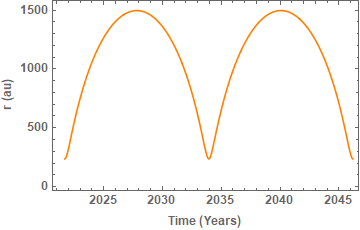}
    \includegraphics[width=7.5cm]{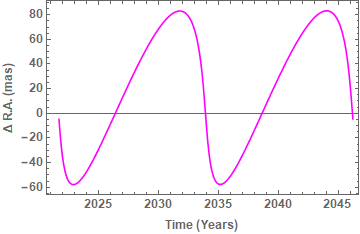}
    \includegraphics[width=7.5cm]{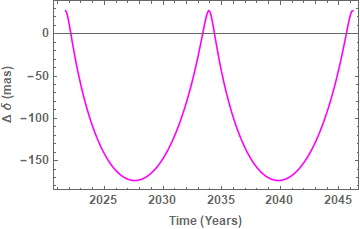}\\

    \includegraphics[width=7.5cm]{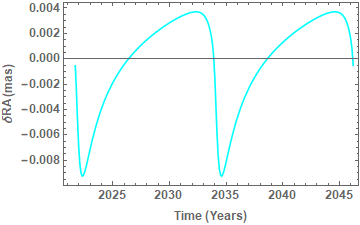}
    \includegraphics[width=7.5cm]{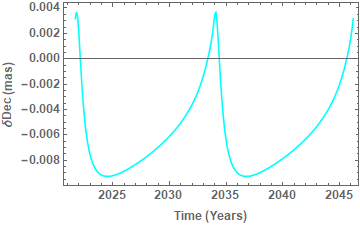}\\
     \caption{Analysis for S55. First row: $\gamma(t)$ (top left) and $r(t)$ (top right)
     as functions of time; 
     second row: right ascension (on the left) and declination (on the right); last row: astrometric shifts in right ascension (left) and declination (on the right). Computations are done starting from $T_0$ over two periods.
     }
    \label{extra55}
\end{figure}

Figs. \ref{extra}, \ref{extra38}, \ref{extra55} show the angular coordinates $\Delta R.A.$ and $\Delta Dec.$ as functions of time for the three stars S2, S38 and S55 respectively. In addition, we include plots of the alignment angle $\gamma$ and the radial distance from the black hole $r$. These are the two quantities that determine the amount of the astrometric shift, which is finally reported in the bottom panels of each figure for both coordinates. We can appreciate that while the angular distances from the black hole are of the order of tens or hundred mas, the astrometric shifts are of the order of ten $\mu$as. Such small astrometric shifts are due to the fact that for all the three stars considered here, the alignment angle $\gamma$ remains quite far from perfect alignment. For S2 we have $\ang{45}<\gamma<\ang{135}$, for S38 $\ang{77}<\gamma<\ang{103}$, for S55 $\ang{69}<\gamma<\ang{111}$.

Table \ref{tabdel} contains the results of our investigation of the systematic shifts in the orbital parameters due to gravitational lensing compared with the statistical uncertainties as declared by GRAVITY. As expected, the estimates of the parameters are robust against gravitational lensing: all systematic errors are well below the uncertainties, but not too far! For S2 we see that for most parameters the shift is just one order of magnitude below the current uncertainty in the parameters. Even keeping the current astrometric accuracy for each individual data point, just by adding new observations in future years it will be possible to let the astrometric shift emerge and be detected. This is indeed an interesting perspective, since the astrometric shift provides another constraint on the enclosed mass that should be compatible with the dynamically derived mass. For S38 and S55 the situation is just slightly less favorable for lensing detection because the orbits are closer to face-on and the periastron is not as close as for S2. For these two stars, the most affected parameter remains the semimajor axis.

\begin{table}[!htbp]
\centering
\begin{tabular}{|l|c|c|c|c|c|c|}
    \hline
\multicolumn{1}{|c|}{} & \multicolumn{2}{c|}{\textbf{S2}} & \multicolumn{2}{c|}{\textbf{S38}} & \multicolumn{2}{c|}{\textbf{S55}}\\
\hline
\textbf{Parameter} & $\boldsymbol{ \sigma (stat.)
}$  & $\boldsymbol{ \delta (syst.)}$ & $\boldsymbol{ \sigma (stat.)
}$  &  $\boldsymbol{ \delta (syst.)}$ & $\boldsymbol{ \sigma (stat.)
}$  &  $\boldsymbol{ \delta (syst.)}$\\
\hline
$a$ (mas) & $4.0\times10^{-2}$ & $3.5\times10^{-3}$ & $4.0\times10^{-2}$ & $3.2\times10^{-3}$  & $5.0\times10^{-2}$ & $5.0\times10^{-3}$\\
$e$  & $6.0\times10^{-5}$ & $4.7\times10^{-6}$ & $1.5\times10^{-4}$ & $6.1\times10^{-6}$ & $2.0\times10^{-4}$ & $7.8\times10^{-6}$\\
$i$  ($^{\circ}$)    &  $3.0\times10^{-2}$ & $-1.2\times10^{-3}$ & $4.0\times10^{-1}$  &  $3.9\times10^{-3}$ & $2.2\times10^{-1}$  &  $4.6\times10^{-3}$\\
$\omega$ ($^{\circ}$) &  $3.0\times10^{-2}$  & $1.9\times10^{-3}$   & $1.0$ &  $2.9\times10^{-3}$   & 
$1.1$ &  $-1.3\times10^{-2}$\\
$\Omega$ ($^{\circ}$)  & $3.0\times10^{-2}$ &  $1.9\times10^{-3}$ &  $1.0$  & $2.9\times10^{-2}$  & 
$1.1$ & $-1.3\times10^{-2}$ \\
$P$  (years)  &   $1.0\times10^{-3}$ & $-2.9\times10^{-5}$ & $1.0\times10^{-2}$ &   $9.9\times10^{-7}$ & $1.0\times10^{-2}$ &   $-5.3\times10^{-6}$  \\
\hline
\end{tabular}
\caption{\label{tabdel} Comparison between the systematic error $\boldsymbol{\delta}$ committed by ignoring gravitational lensing and the statistical uncertainty $\sigma$ reported by the GRAVITY Collaboration \cite{2022A&A...657A..82G,2020A&A...636L...5G} for S2, S38 and S55.}
\end{table}


\section{Exploring Gravitational lensing at various semimajor axes and inclinations}
\label{results}


As shown in the previous section, the astrometric shifts of S2 and the other stars are about one order of magnitude smaller than the current uncertainties in the parameters. Indeed, their orbital inclinations keep the stars relatively far from perfect alignment. For stars getting closer to the black hole, the emitted light also experiences stronger deflections. Motivated by the interest in the detection of effects related to gravitational lensing, in this section we repeat the analysis for S2 after varying the two parameters that have the strongest impact on gravitational lensing: the semimajor axis and the orbital inclination. 

We start from the original value of the semimajor axis $a$ reported in  Table \ref{tab:orpar}, and go down to $a\times10^{-1.5}\approx381$ $R_S$ in steps of $10^{-0.25}$ with a minimum periastron of $44$ $R_S$. The smallest value of $a$ is safely far from the \textit{tidal radius}, i.e. the maximal distance from the MBH where the tidal forces of the MBH would overwhelm the stellar self-gravity and tear the star apart \cite{lodato2015recent}:

\begin{equation}
R_T=R_{S2}\biggl(\frac{M_{BH}}{M_{S2}}\biggr)^{1/3}
\approx 28 R_S \end{equation}
where we have adopted $R_{S2}=8.4$ $R_\odot$ for a normal main sequence star with $M=M_{S2}$ \cite{Eker_2018}.
    
For each of the six semimajor axes in the described range, we also vary the inclination angle $i$ in the range $\ang{90.5}< i< \ang{135}$ in steps of \ang{5}, so as to investigate the effects on the parameter shifts from the S2 inclination $i=\ang{134.7}$ to an edge-on orbit.

In our simulated observations, we keep the number of data points fixed and span two full orbits starting from $T_0$ with a total of 107 data points. Obviously, if the semimajor axis changes, the orbital period changes according to the third Kepler's law.

\begin{figure}[t]
    \centering
    \includegraphics[width=15.5cm]{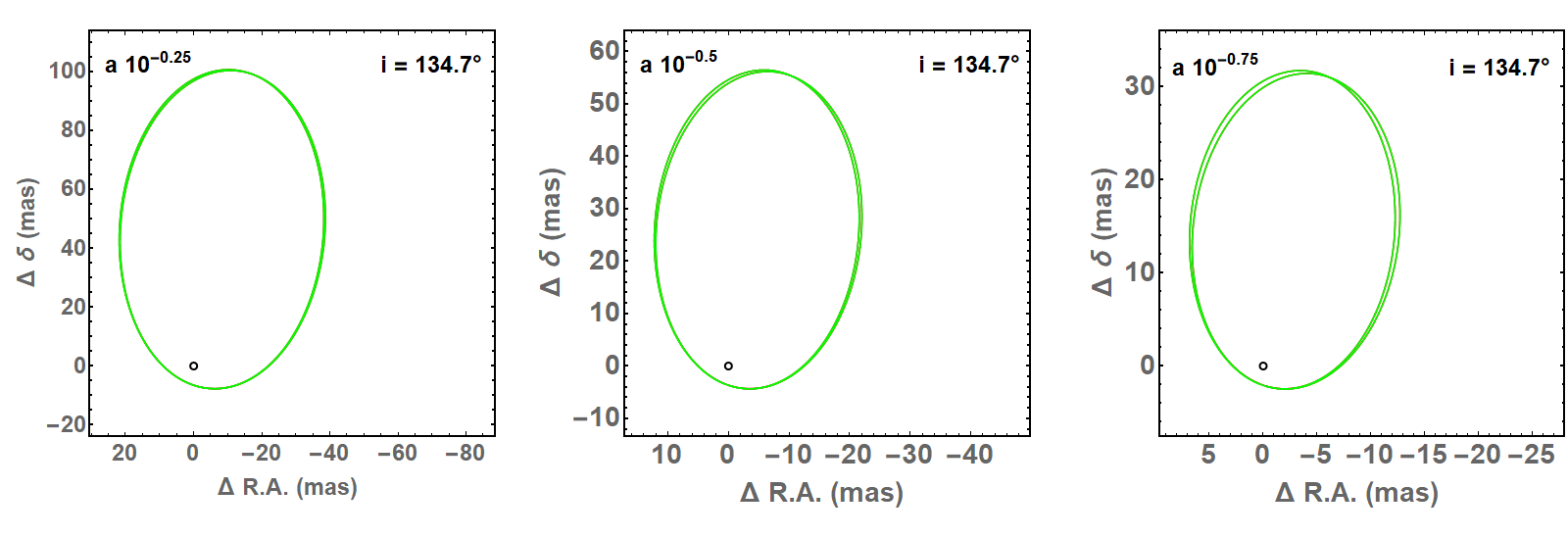}\\
    \includegraphics[width=15.5cm]{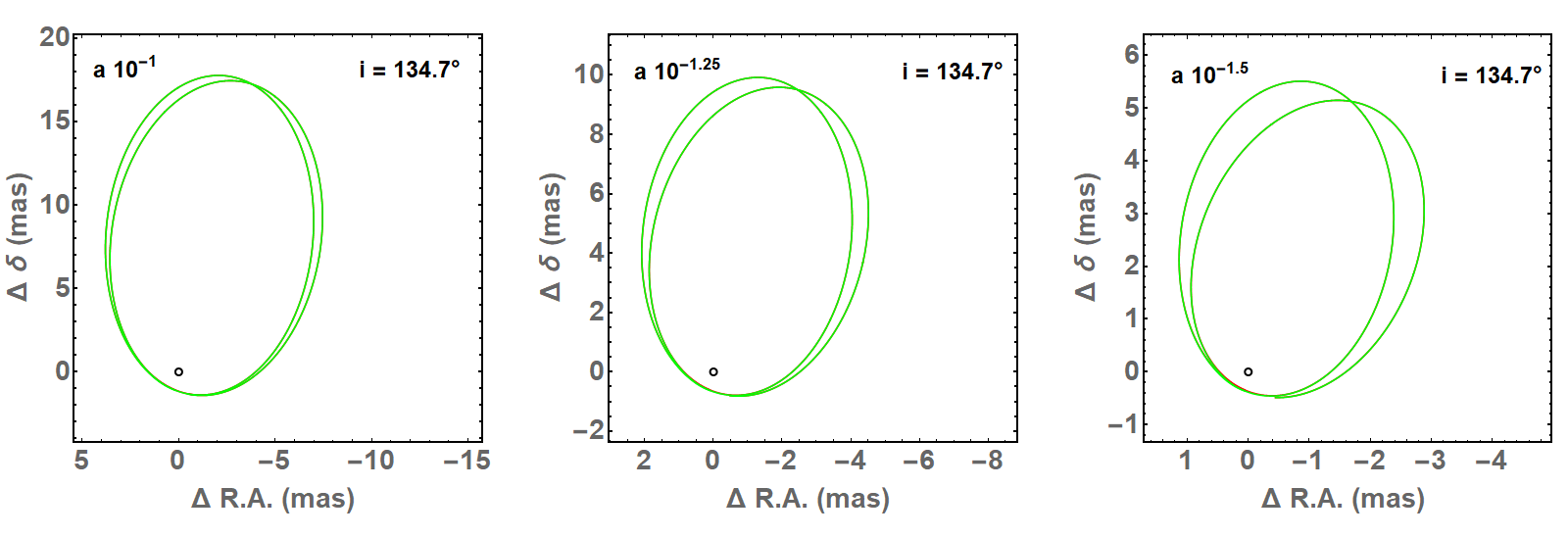}\\
    \caption{S2 orbit with its original fixed inclination $i=\ang{134.7}$ by varying its semimajor axis from $a\times10^{-0.25}$ (first row, top left) to $a\times10^{-1.5}$ (second row, bottom right). The red line stands for the unperturbed orbit, which is almost completely hidden behind the green line standing for the orbit affected by gravitational lensing.
    The circle stands for the position of Sgr A*. The orbit is traced over two periods starting from $T_0$ and it does not close due to the relativistic precession.}
    \label{semimajor}
\end{figure}

In Fig.\ref{semimajor} we show how the S2 orbit changes as the semimajor axis becomes smaller keeping the inclination fixed at the real value. We compare the orbit including the lensing shift (green line) with the orbit without gravitational lensing (red line). The circle marks the position where Sgr A* is located. The orbit is traced over two periods starting from $T_0$ and it does not close due to the relativistic precession. The scale of the figures follows the same factor as the semimajor axis, so as to show the orbits always with the same size. We may note that the green line perfectly overlaps the red line, meaning that the astrometric shift is smaller than the thickness of the lines, save for the figures in the last row, where we can barely appreciate a difference between the green line (including lensing) passing out of the red line (orbit with no lensing). In particular, the effect is visible at the periastron passage in the last panel.

\begin{figure}[t]
    \centering
    \includegraphics[width=16cm]{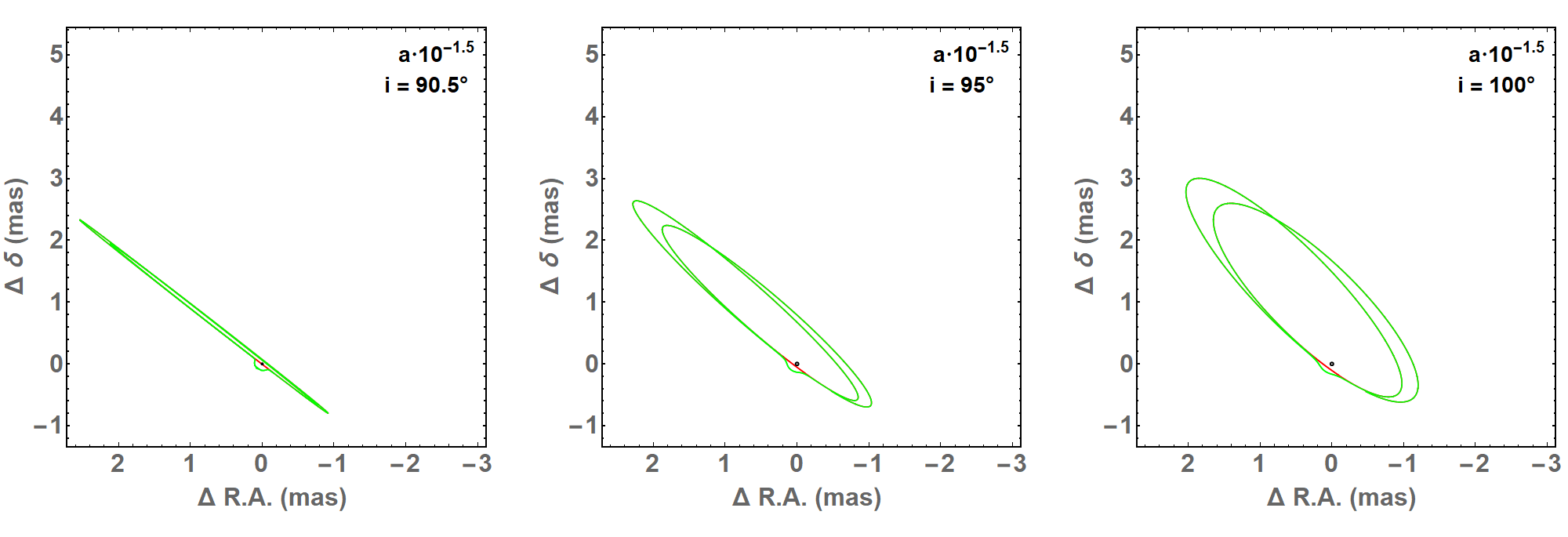}\\
    \includegraphics[width=16cm]{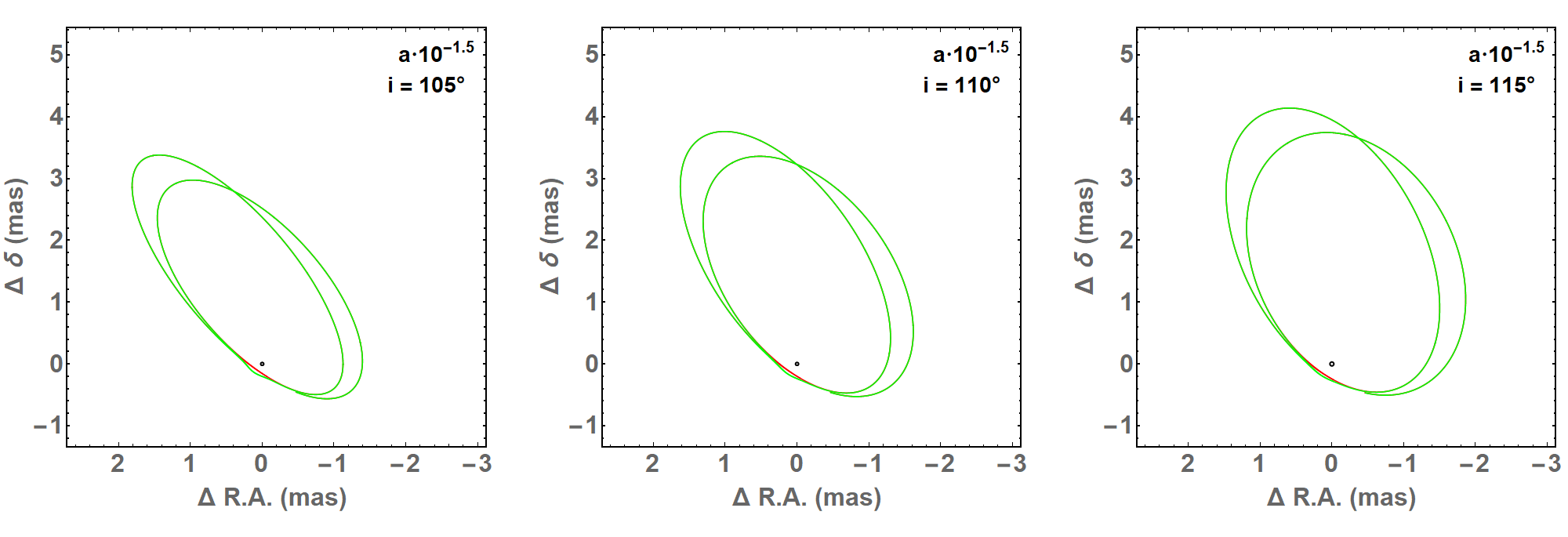}\\
     \includegraphics[width=16cm]{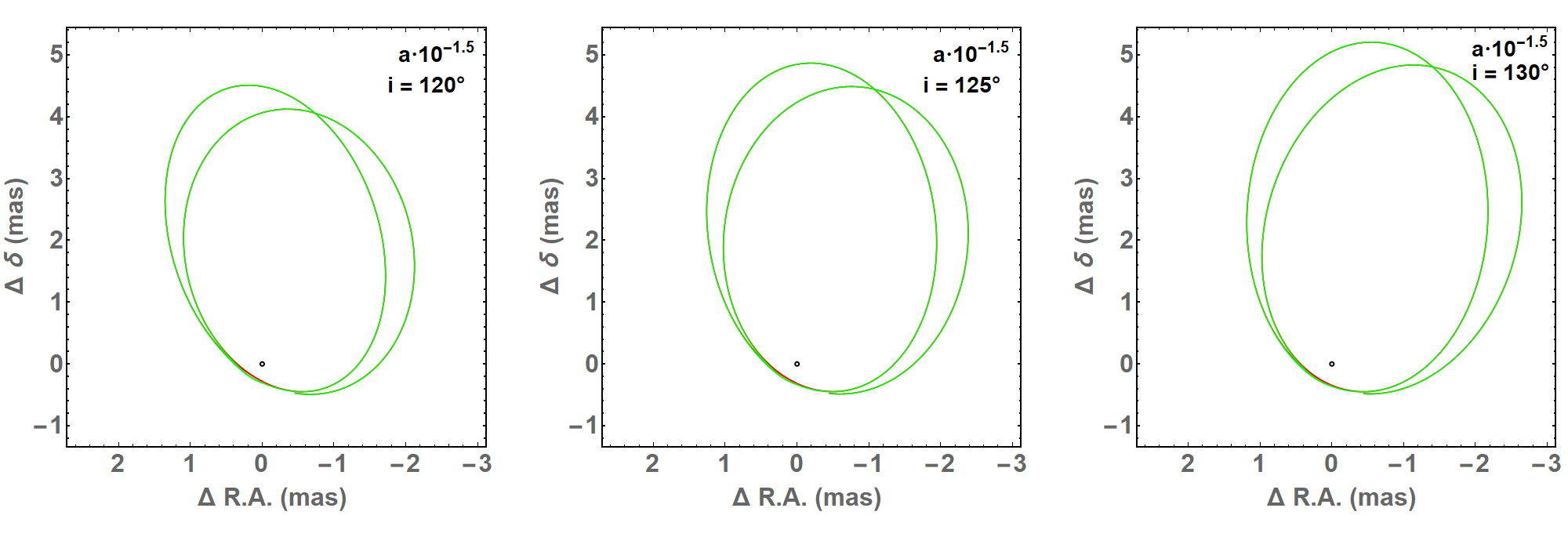}
    \caption{S2 orbit with a fixed reduced semimajor axis by a factor of $10^{-1.5}$  by varying its inclination from $i=\ang{90.5}$ (left top) to $i=\ang{130}$ (right bottom) in steps of \ang{5}. The red line stands for the unperturbed orbit, the green line stands for the orbit affected by gravitational lensing.
    The circle stands for the position of Sgr A*. The orbit is traced over two periods starting from $T_0$ and it does not close due to the relativistic precession.}
    \label{inclination}
\end{figure}

In Fig.\ref{inclination} we show the orbits of S2 with a fixed reduced semimajor axis by a factor of $10^{-1.5}$  and with a varying orbital inclination from $i=\ang{90.5}$ (first row, top left) to $i=\ang{130}$ (second row, bottom right) in steps of \ang{5}. With the smallest semimajor axis, the effect of gravitational lensing is evident in all plots. In particular, nearly edge-on orbits show the typical circular section as the image rapidly revolves around the Einstein ring \cite{1998AcA....48..413J}. 

\begin{figure}[t]
    \centering
\includegraphics[width=15cm]{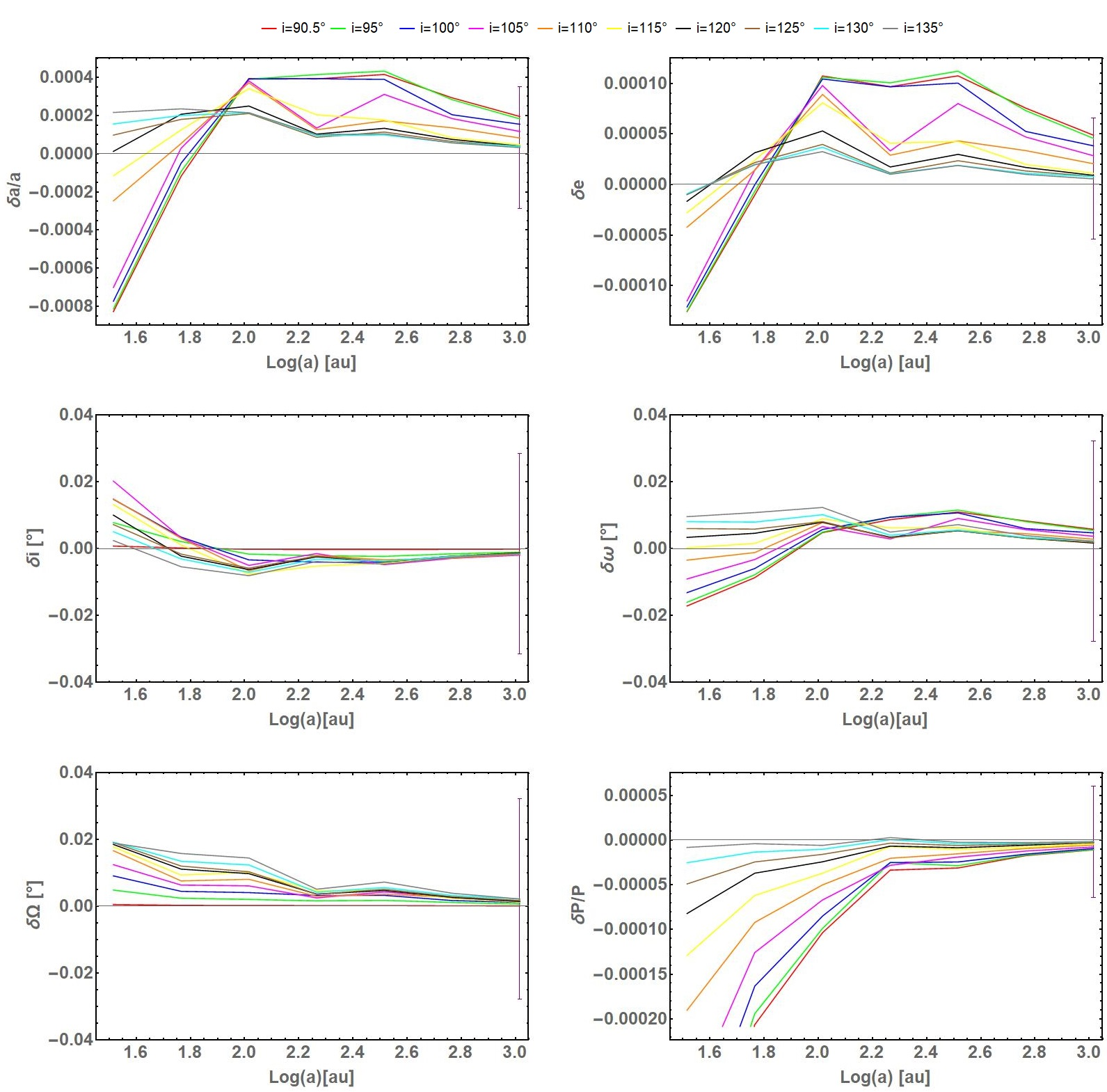}
    \caption{
    The $\boldsymbol{\delta}$ obtained by varying the semimajor axis $a$ and the inclination $i$. From top left to the sixth panel: $\delta_a/a$, $\delta_e$, $\delta_i$, $\delta_\omega$, $\delta_\Omega$, $\delta_P/P$ as a function of $\log a$ expressed in au. 
    The error bars on the right indicate GRAVITY's fit error, see Table \ref{tab:orpar}. 
    }
    \label{deltas}
\end{figure}

Finally, Fig. \ref{deltas} shows the systematic shift $\delta$ in each orbital element due to gravitational lensing for all values of the semimajor axis in our range and for the different inclinations. The maximum semimajor axis corresponds to the actual measured value of S2. Therefore, the gray lines in each plot, which correspond to an inclination of $\ang{135}$, match the values reported in Table \ref{tabdel} at their right ends.

First we note that the shifts in the semimajor axis and the eccentricity are very sensitive to the orbital inclination. For edge-on inclinations, the shifts rapidly approach the current uncertainty in the parameters. If we also decrease the semimajor axis, the shifts exceed the uncertainties. It may look quite counter-intuitive that the shift in the semimajor axis and the eccentricity decreases and even changes sign for closer stars, but we should be aware that the circular deviation of the star image around the Einstein ring cannot be fit by an unlensed orbital trajectory in any effective way. In these cases, the quality of the fit is rapidly degraded and the orbital elements become less and less reliable.

For the angular orbital elements, there is no particular evolution when we go to smaller semimajor axes. For the inclination we note that we start with a negative shift (toward a more edge-on orbit) and we end with a positive shift (more face-on orbit) when the deviation is closer to the Einstein ring. This inversion looks coherent with the plots for the semimajor axis and the eccentricity. A similar effect, but less pronounced, is in the argument of the periastron $\omega$. The shift in the longitude of the ascending node $\Omega$ just increases very slowly.

The negative shift in the period can be interpreted as a combination of the increase in $a$ and the need to follow the relativistic precession: a larger $a$ with the same relativistic precession can only be obtained with a slightly larger mass of the black hole and a slightly shorter period. 

\begin{figure}[t]  

   \centering
    \includegraphics[width=12cm]{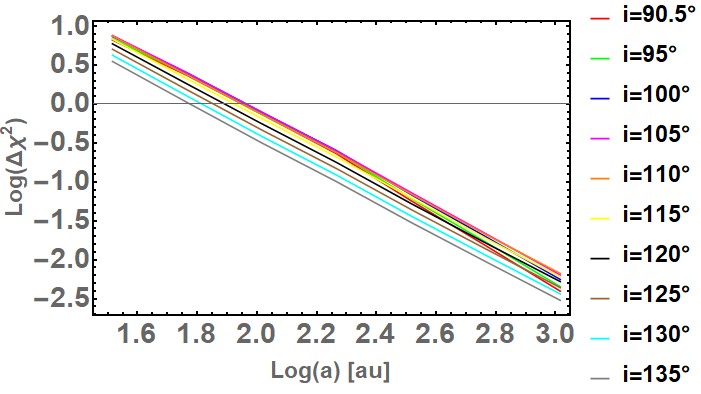}
\caption{The $\chi^2$ test of all data sets vs $\log a$ expressed in au.} \label{Fig chi}   
\end{figure}

Once more we stress the fact that the systematic shifts presented in Fig. \ref{deltas} are independent of the statistical uncertainties. The minimum of the $\chi^2$ does not change if we multiply or divide all $\sigma_i$ by the same factor. Therefore, the significance of the parameter shifts caused by gravitational lensing can be evaluated once a given floor for the statistical noise is assigned. In this respect, Fig. \ref{Fig chi} may be of some help. It represents the change in $\chi^2$ induced by the lensing effects investigated in this section. The $\chi^2$ depends on the uncertainty $\sigma_i$ of the individual measurements and on the total number of data points. In Fig. \ref{Fig chi} we have adopted $\sigma_i=\sigma \left(\frac{a}{a_{act}}\right)$, where $\sigma=0.4$ mas is the average uncertainty of each individual measurement reported by GRAVITY \cite{2022A&A...657A..82G}, $a_{act}$ is the actual semimajor axis as we read it in Table \ref{tab:orpar} and $a$ is the rescaled semimajor axis used in the plots of Fig. \ref{deltas}. In this way, the error bars of each astrometric measurement is rescaled with the same factor as the semimajor axis, which is a reasonable choice for simulated observations in which we assume to be sensitive to smaller and smaller values of $a$. The number of data points in our simulations is fixed to 107. Increasing the data would make the lensing signal emerge by increasing the $\chi^2$. This is particularly true if we add data points in the most sensitive sections of the orbit, when the star is better aligned with the black hole (minima of $\gamma$) and the distance is smaller.

With the basic setup discussed in this section, we see that the $\chi^2$ is very modestly increased until we decrease the semimajor axis by an order of magnitude. Only at this point the $\chi^2$ increases by more than one unit. Actually, we already have thousands of astrometric measurements on S2, which make the detection of lensing effects much closer than what may appear from Fig. \ref{Fig chi}.

\section{Conclusions}
\label{conclusions}

The presence of a MBH at the center of our Galaxy has inspired many studies about light deflection from such a mass concentration surrounded by a very dense stellar environment. Many relativistic effects have already been measured, such as the gravitational redshift, the Shapiro time delay and the relativistic precession of the periapse \cite{2020A&A...636L...5G}. This is indeed happened thanks to the existence of the star S2, which provides an ideal laboratory for testing relativistic dynamics in such strong fields. In contrast with these outstanding measurements, the long sought light deflection effect is surprisingly still hidden within the current astrometric error bars. The only manifestation of light deflection by Sgr A* currently known is through the reconstruction of the image of the shadow by the EHT collaboration \cite{2022ApJ...930L..17E}. However, a clean detection of light deflection on a point source is still missing.

In this paper, we have investigated the impact of gravitational lensing in current orbital fits for S2, S38 and S55. By forcing a fit ignoring lensing on the apparent orbit including the lensing effect, we estimate that the systematic shift in the orbital elements is about one tenth of the reported uncertainties. Therefore, we recommend the inclusion of the astrometric deflection of the image of S2 and other stars in the reconstruction of the orbits already at the present stage. Furthermore, by a continuous follow-up of the orbits of these stars, the light deflection should be detectable even at the current astrometric precision. In particular, it is crucial to have very accurate observations in the most sensitive sections of the orbit, namely when the star has the most favorable alignment with the black hole.

In addition to this analysis focusing on the actual orbits of the S-stars, we have presented an investigation of the systematic shift in the orbital elements after varying the semimajor axis and the orbital inclination of S2. For nearly edge-on orbits the shift in the parameters may increase by an order of magnitude. The effect becomes even more pronounced for stars orbiting closer and closer to the MBH.

\appendix

\section{Orbital
parameters of the real and apparent orbits
}\label{orbpar}

We deal with a two-body problem where S2 and the other S-stars have an elliptic orbit and Sgr A* lies in the focus located at the center of mass and we denote $OXYZ$ the coordinate system of the apparent orbit of the star: the axes origin is centered at Sgr A*. The Z-axis of the coordinate system is defined by the vector pointing from the Solar System to the Galactic Center, and the X and Y axes are defined such that the X-Y plane is parallel to the plane of the sky, with the X-axis pointing East and the Y-axis pointing North; we denote $Oxyz$ the plane of the real star's orbit.

In polar coordinates the ellipse equation in the $Oxyz$ system is given by

\begin{equation}
    \begin{cases}
x=r\cos{\phi}\\y=r\sin{\phi  }\\z=0
    \end{cases}
\end{equation}


where $r(\tau)$ is the star position that we get as a function of the proper time from Eq.\eqref{eqmo} and, in our case, it coincides with $D_{LS}$, i.e. the distance between the lens (Sgr A*) and the source (the S-star); $\phi$ is the true anomaly that we get, as a function of the proper time $\phi(\tau)$, from Eq.\eqref{fipunto}, $a$ is the semimajor axes of the ellipse and $e$ its eccentricity.

We can always obtain the apparent orbit in the $OXYZ$ system by applying a matrix transformation on the $Oxyz$ system:

\begin{equation}
  \begin{pmatrix}
X\\
Y\\
Z
\end{pmatrix}=  R \begin{pmatrix}
x\\
y\\
z
\end{pmatrix}
\end{equation}

where $R=R_\Omega\cdot R_i\cdot R_\omega$ is the rotation matrix where $i$ is the angle of inclination between the real orbit and the observation plane, $\Omega$ is the angle of the ascending node, and  $\omega$ is the argument of pericenter, and

\begin{equation}
 R_\Omega(Z)= \begin{pmatrix}
\cos{\Omega} &&& -\sin{\Omega} &&& 0 \\
\sin{\Omega} &&&  \cos{\Omega} &&& 0\\
0 &&& 0 &&& 1
\end{pmatrix}
\end{equation}

\begin{equation}
 R_i(X)= \begin{pmatrix}
 1 &&& 0 &&& 0\\
0 &&& \cos{i} &&& -\sin{i}\\
0 &&& \sin{i} &&& \cos{i}
\end{pmatrix}
\end{equation}

\begin{equation}
 R_\omega(Y)= \begin{pmatrix}
\cos{\omega} &&& -\sin{\omega} &&& 0 \\
\sin{\omega} &&&  \cos{\omega} &&& 0\\
0 &&& 0 &&& 1
\end{pmatrix}
\end{equation}

where $R_\Omega$ represents the rotation along the Z-axes, $R_i$ the rotation along the X-axes and $R_\omega$ is the rotation along the Y-axes. Then
finally we get the Cartesian coordinates of the apparent orbit of the star on the sky plane, 

\begin{equation}
    \begin{cases}
$$X=r [\cos{\Omega}\cos{(\phi+\omega)}-\sin{\Omega}\sin{(\phi+\omega)} \cos{i}]$$\\ 
$$Y=r [\sin{\Omega}\cos{(\phi+\omega)}+ \cos{\Omega}\sin{(\phi+\omega)}\cos{i}]$$\\
$$Z=r[\sin{(\phi+\omega)}\sin{i}]$$
\end{cases}\label{sys}
\end{equation}


\begin{figure}[!htbp]
\centering 
\includegraphics[width=12cm]{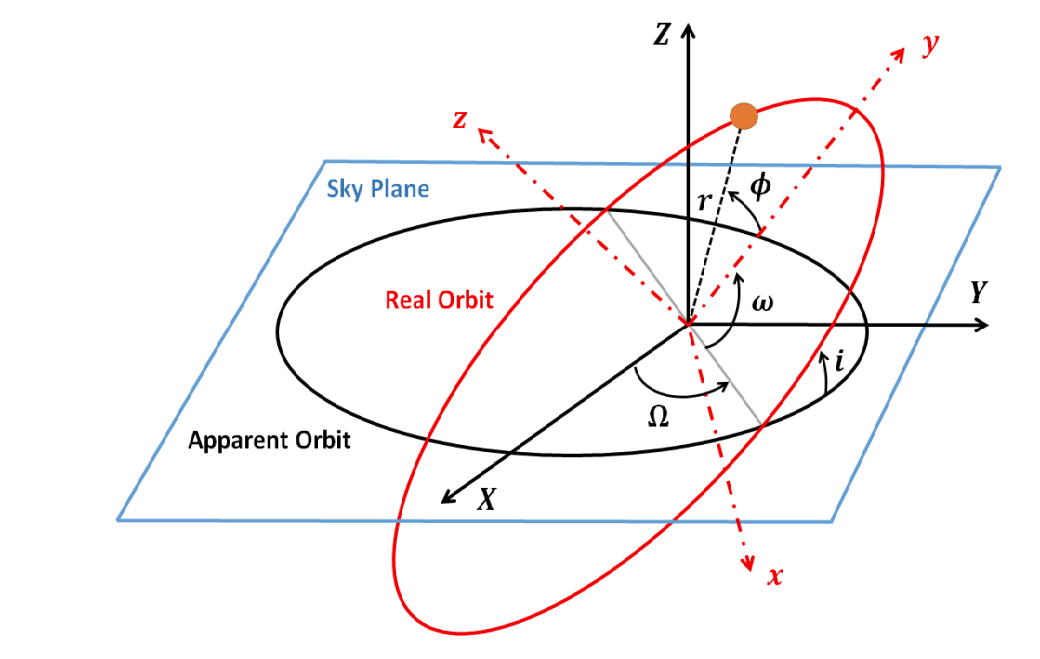}
\hfill
\caption{\label{pianoorb} Projection on the sky of the real star's orbit onto the plane of the sky as in ref. \cite{BecerraVergara2020}, the osculating orbital parameters are shown.}
\label{elor}
\end{figure}

The observed displacement from Sgr A* are 
\begin{equation}
    \begin{cases}
    \Delta Dec. \equiv \frac{X}{D_\bullet} \\
    \Delta R.A. \equiv \frac{Y}{D_\bullet}
    \end{cases}\label{the}.
\end{equation}
When we turn on the gravitational lensing effect, both coordinates are multiplied by $D_\bullet \theta/\sqrt{X^2 + Y^2}$ with $\theta$ given by Eq. (\ref{imp}).

\clearpage

\bibliographystyle{stile.bst}  
\bibliography{mybib.bib}  

\providecommand{\href}[2]{#2}\begingroup\raggedright\begin{thebibliography}{10}

\bibitem{2019Msngr.178...26G}
{GRAVITY Collaboration}, R.~{Abuter}, M.~{Accardo}, T.~{Adler}, A.~{Amorim},
  N.~{Anugu} et~al., \emph{{GRAVITY and the Galactic Centre}},
  \href{https://doi.org/10.18727/0722-6691/5168}{\emph{The Messenger}
  {\bfseries 178} (2019) 26}.

\bibitem{gillessen2009orbit}
S.~Gillessen, F.~Eisenhauer, T.~Fritz, H.~Bartko, K.~Dodds-Eden, O.~Pfuhl
  et~al., \emph{The orbit of the star s2 around sgr a* from very large
  telescope and keck data}, {\emph{The Astrophysical Journal} {\bfseries 707}
  (2009) L114}.

\bibitem{gillessen2017update}
S.~Gillessen, P.~Plewa, F.~Eisenhauer, R.~Sari, I.~Waisberg, M.~Habibi et~al.,
  \emph{An update on monitoring stellar orbits in the galactic center},
  {\emph{The Astrophysical Journal} {\bfseries 837} (2017) 30}.

\bibitem{2013CQGra..30x4003F}
H.~{Falcke} and S.B.~{Markoff}, \emph{{Toward the event horizon{\textemdash}the
  supermassive black hole in the Galactic Center}},
  \href{https://doi.org/10.1088/0264-9381/30/24/244003}{\emph{Classical and
  Quantum Gravity} {\bfseries 30} (2013) 244003}
  [\href{https://arxiv.org/abs/1311.1841}{{\ttfamily 1311.1841}}].

\bibitem{2003ApJ...586L.127G}
A.M.~{Ghez}, G.~{Duch{\^e}ne}, K.~{Matthews}, S.D.~{Hornstein}, A.~{Tanner},
  J.~{Larkin} et~al., \emph{{The First Measurement of Spectral Lines in a
  Short-Period Star Bound to the Galaxy's Central Black Hole: A Paradox of
  Youth}}, \href{https://doi.org/10.1086/374804}{\emph{ApJL} {\bfseries 586}
  (2003) L127} [\href{https://arxiv.org/abs/astro-ph/0302299}{{\ttfamily
  astro-ph/0302299}}].

\bibitem{2005ApJ...628..246E}
F.~{Eisenhauer}, R.~{Genzel}, T.~{Alexander}, R.~{Abuter}, T.~{Paumard},
  T.~{Ott} et~al., \emph{{SINFONI in the Galactic Center: Young Stars and
  Infrared Flares in the Central Light-Month}},
  \href{https://doi.org/10.1086/430667}{\emph{ApJ} {\bfseries 628} (2005) 246}
  [\href{https://arxiv.org/abs/astro-ph/0502129}{{\ttfamily
  astro-ph/0502129}}].

\bibitem{2009ApJ...692.1075G}
S.~{Gillessen}, F.~{Eisenhauer}, S.~{Trippe}, T.~{Alexander}, R.~{Genzel},
  F.~{Martins} et~al., \emph{{Monitoring Stellar Orbits Around the Massive
  Black Hole in the Galactic Center}},
  \href{https://doi.org/10.1088/0004-637X/692/2/1075}{\emph{ApJ} {\bfseries
  692} (2009) 1075} [\href{https://arxiv.org/abs/0810.4674}{{\ttfamily
  0810.4674}}].

\bibitem{2017ApJ...837...30G}
S.~{Gillessen}, P.M.~{Plewa}, F.~{Eisenhauer}, R.~{Sari}, I.~{Waisberg},
  M.~{Habibi} et~al., \emph{{An Update on Monitoring Stellar Orbits in the
  Galactic Center}}, \href{https://doi.org/10.3847/1538-4357/aa5c41}{\emph{ApJ}
  {\bfseries 837} (2017) 30}
  [\href{https://arxiv.org/abs/1611.09144}{{\ttfamily 1611.09144}}].

\bibitem{2012Sci...338...84M}
L.~{Meyer}, A.M.~{Ghez}, R.~{Sch{\"o}del}, S.~{Yelda}, A.~{Boehle}, J.R.~{Lu}
  et~al., \emph{{The Shortest-Known-Period Star Orbiting Our
  Galaxy{\textquoteright}s Supermassive Black Hole}},
  \href{https://doi.org/10.1126/science.1225506}{\emph{Science} {\bfseries 338}
  (2012) 84} [\href{https://arxiv.org/abs/1210.1294}{{\ttfamily 1210.1294}}].

\bibitem{2016ApJ...830...17B}
A.~{Boehle}, A.M.~{Ghez}, R.~{Sch{\"o}del}, L.~{Meyer}, S.~{Yelda}, S.~{Albers}
  et~al., \emph{{An Improved Distance and Mass Estimate for Sgr A* from a
  Multistar Orbit Analysis}},
  \href{https://doi.org/10.3847/0004-637X/830/1/17}{\emph{ApJ} {\bfseries 830}
  (2016) 17} [\href{https://arxiv.org/abs/1607.05726}{{\ttfamily 1607.05726}}].

\bibitem{2016ApJ...821...44F}
T.K.~{Fritz}, S.~{Chatzopoulos}, O.~{Gerhard}, S.~{Gillessen}, R.~{Genzel},
  O.~{Pfuhl} et~al., \emph{{The Nuclear Cluster of the Milky Way: Total Mass
  and Luminosity}},
  \href{https://doi.org/10.3847/0004-637X/821/1/44}{\emph{ApJ} {\bfseries 821}
  (2016) 44} [\href{https://arxiv.org/abs/1406.7568}{{\ttfamily 1406.7568}}].

\bibitem{2002Natur.419..694S}
R.~{Sch{\"o}del}, T.~{Ott}, R.~{Genzel}, R.~{Hofmann}, M.~{Lehnert},
  A.~{Eckart} et~al., \emph{{A star in a 15.2-year orbit around the
  supermassive black hole at the centre of the Milky Way}},
  \href{https://doi.org/10.1038/nature01121}{\emph{Nature} {\bfseries 419}
  (2002) 694} [\href{https://arxiv.org/abs/astro-ph/0210426}{{\ttfamily
  astro-ph/0210426}}].

\bibitem{2012RAA....12..995M}
M.R.~{Morris}, L.~{Meyer} and A.M.~{Ghez}, \emph{{Galactic center research:
  manifestations of the central black hole}},
  \href{https://doi.org/10.1088/1674-4527/12/8/007}{\emph{Research in Astronomy
  and Astrophysics} {\bfseries 12} (2012) 995}
  [\href{https://arxiv.org/abs/1207.6755}{{\ttfamily 1207.6755}}].

\bibitem{2009A&A...502...91S}
R.~{Sch{\"o}del}, D.~{Merritt} and A.~{Eckart}, \emph{{The nuclear star cluster
  of the Milky Way: proper motions and mass}},
  \href{https://doi.org/10.1051/0004-6361/200810922}{\emph{Astronomy and
  Astrophysics} {\bfseries 502} (2009) 91}
  [\href{https://arxiv.org/abs/0902.3892}{{\ttfamily 0902.3892}}].

\bibitem{2021A&A...645A.127G}
{Gravity Collaboration}, R.~{Abuter}, A.~{Amorim}, M.~{Baub{\"o}ck},
  J.P.~{Berger}, H.~{Bonnet} et~al., \emph{{Detection of faint stars near
  Sagittarius A* with GRAVITY}},
  \href{https://doi.org/10.1051/0004-6361/202039544}{\emph{Astronomy and
  Astrophysics} {\bfseries 645} (2021) A127}
  [\href{https://arxiv.org/abs/2011.03058}{{\ttfamily 2011.03058}}].

\bibitem{2008SPIE.7013E..2AE}
F.~{Eisenhauer}, G.~{Perrin}, W.~{Brandner}, C.~{Straubmeier}, A.~{Richichi},
  S.~{Gillessen} et~al., \emph{{GRAVITY: getting to the event horizon of Sgr
  A*}},  in \emph{Optical and Infrared Interferometry}, M.~{Sch{\"o}ller},
  W.C.~{Danchi} and F.~{Delplancke}, eds., vol.~7013 of \emph{Society of
  Photo-Optical Instrumentation Engineers (SPIE) Conference Series}, p.~70132A,
  July, 2008, \href{https://doi.org/10.1117/12.788407}{DOI}
  [\href{https://arxiv.org/abs/0808.0063}{{\ttfamily 0808.0063}}].

\bibitem{2019hsax.conf..609A}
R.~{Abuter}, A.~{Amorim}, N.~{Anugu}, M.~{Baub{\"o}ck}, M.~{Benisty},
  J.P.~{Berger} et~al., \emph{{GRAVITY - Reaching out to SgrA* with VLTI}},  in
  \emph{Highlights on Spanish Astrophysics X}, B.~{Montesinos}, A.~{Asensio
  Ramos}, F.~{Buitrago}, R.~{Sch{\"o}del}, E.~{Villaver}, S.~{P{\'e}rez-Hoyos}
  et~al., eds., pp.~609--610, Mar., 2019.

\bibitem{2021A&A...647A..59G}
{Gravity Collaboration}, R.~{Abuter}, A.~{Amorim}, M.~{Baub{\"o}ck},
  J.P.~{Berger}, H.~{Bonnet} et~al., \emph{{Improved GRAVITY astrometric
  accuracy from modeling optical aberrations}},
  \href{https://doi.org/10.1051/0004-6361/202040208}{\emph{Astronomy and
  Astrophysics} {\bfseries 647} (2021) A59}
  [\href{https://arxiv.org/abs/2101.12098}{{\ttfamily 2101.12098}}].

\bibitem{2017A&A...602A..94G}
{Gravity Collaboration}, R.~{Abuter}, M.~{Accardo}, A.~{Amorim}, N.~{Anugu},
  G.~{{\'A}vila} et~al., \emph{{First light for GRAVITY: Phase referencing
  optical interferometry for the Very Large Telescope Interferometer}},
  \href{https://doi.org/10.1051/0004-6361/201730838}{\emph{Astronomy and
  Astrophysics} {\bfseries 602} (2017) A94}
  [\href{https://arxiv.org/abs/1705.02345}{{\ttfamily 1705.02345}}].

\bibitem{2020A&A...636L...5G}
{Gravity Collaboration}, R.~{Abuter}, A.~{Amorim}, M.~{Baub{\"o}ck},
  J.P.~{Berger}, H.~{Bonnet} et~al., \emph{{Detection of the Schwarzschild
  precession in the orbit of the star S2 near the Galactic centre massive black
  hole}}, \href{https://doi.org/10.1051/0004-6361/202037813}{\emph{Astronomy
  and Astrophysics} {\bfseries 636} (2020) L5}
  [\href{https://arxiv.org/abs/2004.07187}{{\ttfamily 2004.07187}}].

\bibitem{1992ApJ...387L..65W}
M.~{Wardle} and F.~{Yusef-Zadeh}, \emph{{Gravitational Lensing by a Massive
  Black Hole at the Galactic Center}},
  \href{https://doi.org/10.1086/186306}{\emph{The Astrophysical Journal
  Letters} {\bfseries 387} (1992) L65}.

\bibitem{1998AcA....48..413J}
M.~{Jaroszynski}, \emph{{Gravitational Lensing and Proper Motions of Stars
  Surrounding the Galactic Center}}, {\emph{Acta Astronomica} {\bfseries 48}
  (1998) 413}.

\bibitem{2003A&A...409..809D}
F.~{De Paolis}, A.~{Geralico}, G.~{Ingrosso} and A.A.~{Nucita}, \emph{{The
  black hole at the galactic center as a possible retro-lens for the S2
  orbiting star}},
  \href{https://doi.org/10.1051/0004-6361:20031137}{\emph{Astronomy and
  Astrophysics} {\bfseries 409} (2003) 809}
  [\href{https://arxiv.org/abs/astro-ph/0307493}{{\ttfamily
  astro-ph/0307493}}].

\bibitem{2004ApJ...611.1045B}
V.~{Bozza} and L.~{Mancini}, \emph{{Gravitational Lensing by Black Holes: A
  Comprehensive Treatment and the Case of the Star S2}},
  \href{https://doi.org/10.1086/422309}{\emph{ApJ} {\bfseries 611} (2004) 1045}
  [\href{https://arxiv.org/abs/astro-ph/0404526}{{\ttfamily
  astro-ph/0404526}}].

\bibitem{2005ApJ...627..790B}
V.~{Bozza} and L.~{Mancini}, \emph{{Gravitational Lensing of Stars in the
  Central Arcsecond of Our Galaxy}},
  \href{https://doi.org/10.1086/430664}{\emph{ApJ} {\bfseries 627} (2005) 790}
  [\href{https://arxiv.org/abs/astro-ph/0503664}{{\ttfamily
  astro-ph/0503664}}].

\bibitem{2021ApJ...915L..33M}
M.J.~{Micha{\l}owski} and P.~{Mr{\'o}z}, \emph{{Stars Lensed by the
  Supermassive Black Hole in the Center of the Milky Way: Predictions for ELT,
  TMT, GMT, and JWST}},
  \href{https://doi.org/10.3847/2041-8213/ac0f81}{\emph{The Astrophysical
  Journal Letters} {\bfseries 915} (2021) L33}
  [\href{https://arxiv.org/abs/2107.00659}{{\ttfamily 2107.00659}}].

\bibitem{2012ApJ...753...56B}
V.~{Bozza} and L.~{Mancini}, \emph{{Observing Gravitational Lensing Effects by
  Sgr A* with GRAVITY}},
  \href{https://doi.org/10.1088/0004-637X/753/1/56}{\emph{Astrophysical
  Journal} {\bfseries 753} (2012) 56}
  [\href{https://arxiv.org/abs/1204.2103}{{\ttfamily 1204.2103}}].

\bibitem{2015ApJ...809..127Z}
F.~{Zhang}, Y.~{Lu} and Q.~{Yu}, \emph{{On Testing the Kerr Metric of the
  Massive Black Hole in the Galactic Center via Stellar Orbital Motion: Full
  General Relativistic Treatment}},
  \href{https://doi.org/10.1088/0004-637X/809/2/127}{\emph{Astrophysical
  Journal} {\bfseries 809} (2015) 127}
  [\href{https://arxiv.org/abs/1508.06293}{{\ttfamily 1508.06293}}].

\bibitem{2017A&A...608A..60G}
M.~{Grould}, F.H.~{Vincent}, T.~{Paumard} and G.~{Perrin}, \emph{{General
  relativistic effects on the orbit of the S2 star with GRAVITY}},
  \href{https://doi.org/10.1051/0004-6361/201731148}{\emph{Astronomy and
  Astrophysics} {\bfseries 608} (2017) A60}
  [\href{https://arxiv.org/abs/1709.04492}{{\ttfamily 1709.04492}}].

\bibitem{2018MNRAS.476.3600W}
I.~{Waisberg}, J.~{Dexter}, S.~{Gillessen}, O.~{Pfuhl}, F.~{Eisenhauer},
  P.M.~{Plewa} et~al., \emph{{What stellar orbit is needed to measure the spin
  of the Galactic centre black hole from astrometric data?}},
  \href{https://doi.org/10.1093/mnras/sty476}{\emph{Monthly Notices of the RAS}
  {\bfseries 476} (2018) 3600}
  [\href{https://arxiv.org/abs/1802.08198}{{\ttfamily 1802.08198}}].

\bibitem{2018A&A...609A..28B}
H.~{Baumgardt}, P.~{Amaro-Seoane} and R.~{Sch{\"o}del}, \emph{{The distribution
  of stars around the Milky Way's central black hole. III. Comparison with
  simulations}},
  \href{https://doi.org/10.1051/0004-6361/201730462}{\emph{Astronomy and
  Astrophysics} {\bfseries 609} (2018) A28}
  [\href{https://arxiv.org/abs/1701.03818}{{\ttfamily 1701.03818}}].

\bibitem{2008ApJ...672L.119M}
F.~{Martins}, S.~{Gillessen}, F.~{Eisenhauer}, R.~{Genzel}, T.~{Ott} and
  S.~{Trippe}, \emph{{On the Nature of the Fast-Moving Star S2 in the Galactic
  Center}}, \href{https://doi.org/10.1086/526768}{\emph{ApJL} {\bfseries 672}
  (2008) L119} [\href{https://arxiv.org/abs/0711.3344}{{\ttfamily 0711.3344}}].

\bibitem{2010MNRAS.409.1146G}
A.~{Gualandris}, S.~{Gillessen} and D.~{Merritt}, \emph{{The Galactic Centre
  star S2 as a dynamical probe for intermediate-mass black holes}},
  \href{https://doi.org/10.1111/j.1365-2966.2010.17373.x}{\emph{Monthly Notices
  of the RAS} {\bfseries 409} (2010) 1146}
  [\href{https://arxiv.org/abs/1006.3563}{{\ttfamily 1006.3563}}].

\bibitem{2004PhyA..332...89C}
P.-H.~{Chavanis}, \emph{{Generalized thermodynamics and kinetic equations:
  Boltzmann, Landau, Kramers and Smoluchowski}},
  \href{https://doi.org/10.1016/j.physa.2003.09.061}{\emph{Physica A
  Statistical Mechanics and its Applications} {\bfseries 332} (2004) 89}
  [\href{https://arxiv.org/abs/cond-mat/0304073}{{\ttfamily
  cond-mat/0304073}}].

\bibitem{2017ApJ...847..120H}
M.~{Habibi}, S.~{Gillessen}, F.~{Martins}, F.~{Eisenhauer}, P.M.~{Plewa},
  O.~{Pfuhl} et~al., \emph{{Twelve Years of Spectroscopic Monitoring in the
  Galactic Center: The Closest Look at S-stars near the Black Hole}},
  \href{https://doi.org/10.3847/1538-4357/aa876f}{\emph{ApJ} {\bfseries 847}
  (2017) 120} [\href{https://arxiv.org/abs/1708.06353}{{\ttfamily
  1708.06353}}].

\bibitem{2022A&A...657A..82G}
{Gravity Collaboration}, R.~{Abuter}, N.~{Aimar}, A.~{Amorim}, P.~{Arras},
  M.~{Baub{\"o}ck} et~al., \emph{{Deep images of the Galactic center with
  GRAVITY}}, \href{https://doi.org/10.1051/0004-6361/202142459}{\emph{Astronomy
  and Astrophysics} {\bfseries 657} (2022) A82}
  [\href{https://arxiv.org/abs/2112.07477}{{\ttfamily 2112.07477}}].

\bibitem{2009ApJ...707L.114G}
S.~{Gillessen}, F.~{Eisenhauer}, T.K.~{Fritz}, H.~{Bartko}, K.~{Dodds-Eden},
  O.~{Pfuhl} et~al., \emph{{The Orbit of the Star S2 Around SGR A* from Very
  Large Telescope and Keck Data}},
  \href{https://doi.org/10.1088/0004-637X/707/2/L114}{\emph{ApJL} {\bfseries
  707} (2009) L114} [\href{https://arxiv.org/abs/0910.3069}{{\ttfamily
  0910.3069}}].

\bibitem{2018A&A...618L..10G}
{Gravity Collaboration}, R.~{Abuter}, A.~{Amorim}, M.~{Baub{\"o}ck},
  J.P.~{Berger}, H.~{Bonnet} et~al., \emph{{Detection of orbital motions near
  the last stable circular orbit of the massive black hole SgrA*}},
  \href{https://doi.org/10.1051/0004-6361/201834294}{\emph{Astronomy and
  Astrophysics} {\bfseries 618} (2018) L10}
  [\href{https://arxiv.org/abs/1810.12641}{{\ttfamily 1810.12641}}].

\bibitem{BecerraVergara2020}
E.A.~Becerra-Vergara, C.R.~Argüelles, A.~Krut, J.A.~Rueda and R.~Ruffini,
  \emph{Geodesic motion of s2 and g2 as a test of the fermionic dark matter
  nature of our galactic core},
  \href{https://doi.org/10.1051/0004-6361/201935990}{\emph{Astronomy {\&}
  Astrophysics} {\bfseries 641} (2020) A34}.

\bibitem{2019Sci...365..664D}
T.~{Do}, A.~{Hees}, A.~{Ghez}, G.D.~{Martinez}, D.S.~{Chu}, S.~{Jia} et~al.,
  \emph{{Relativistic redshift of the star S0-2 orbiting the Galactic Center
  supermassive black hole}},
  \href{https://doi.org/10.1126/science.aav8137}{\emph{Science} {\bfseries 365}
  (2019) 664} [\href{https://arxiv.org/abs/1907.10731}{{\ttfamily
  1907.10731}}].

\bibitem{madsen2004methods}
K.~Madsen, H.B.~Nielsen and O.~Tingleff, \emph{Methods for non-linear least
  squares problems}, .

\bibitem{lodato2015recent}
G.~Lodato, A.~Franchini, C.~Bonnerot and E.M.~Rossi, \emph{Recent developments
  in the theory of tidal disruption events}, {\emph{Journal of High Energy
  Astrophysics} {\bfseries 7} (2015) 158}.

\bibitem{Eker_2018}
Z.~Eker, V.~Bak{\i}{\c{s} }, S.~Bilir, F.~Soydugan, I.~Steer, E.~Soydugan
  et~al., \emph{Interrelated main-sequence mass{\textendash}luminosity,
  mass{\textendash}radius, and mass{\textendash}effective temperature
  relations}, \href{https://doi.org/10.1093/mnras/sty1834}{\emph{Monthly
  Notices of the Royal Astronomical Society} {\bfseries 479} (2018) 5491}.

\bibitem{2022ApJ...930L..17E}
{Event Horizon Telescope Collaboration}, K.~{Akiyama}, A.~{Alberdi}, W.~{Alef},
  J.C.~{Algaba}, R.~{Anantua} et~al., \emph{{First Sagittarius A* Event Horizon
  Telescope Results. VI. Testing the Black Hole Metric}},
  \href{https://doi.org/10.3847/2041-8213/ac6756}{\emph{The Astrophysical
  Journal Letters} {\bfseries 930} (2022) L17}.

\end{thebibliography}\endgroup

\end{document}